\documentclass[conference]{IEEEtran}

\usepackage{amsmath}
\usepackage{amssymb}
\usepackage{amsfonts}

\usepackage{graphicx}
\usepackage{comment}

\usepackage{booktabs}
\usepackage{multirow}
\usepackage{array}

\usepackage{cite}

\usepackage{url}

\usepackage{xcolor}

\usepackage[hidelinks]{hyperref}

\title{EPR Count for Runtime Prediction in Distributed Quantum Computing}

\author{
    Fatih E. Bilgen$^{*}$, Ozgur B. Akan$^{*\dagger}$\\

    $^{*}$Centre for neXt Communications (CXC), Electrical Engineering Division, Department of Engineering,\\
    University of Cambridge, Cambridge, UK\\

    $^{\dagger}$Centre for neXt Communications (CXC), Department of Electrical and Electronics Engineering,\\
    Ko\c{c} University, Istanbul, Turkey\\

    Email: \{feb49, oba21\}@cam.ac.uk
}

\begin{document}

\maketitle


\begin{abstract}
EPR-pair consumption is commonly used as a communication-cost objective in distributed quantum computing, but minimizing EPR cost does not necessarily minimize distributed execution time. Despite its widespread use, the reliability of EPR count as a runtime surrogate has received limited direct characterization across different workloads and communication conditions. This work addresses this gap by systematically evaluating the relationship between EPR cost and runtime over all balanced two-QPU mappings of QFT, QAOA, and CDKM circuits. The results show that mappings with identical EPR cost can have substantially different execution times, lower-EPR mappings can be slower, and minimum-EPR mappings can be runtime-suboptimal. The reliability of EPR count is strongly workload dependent and also changes with the communication operating regime: increasing EPR-generation latency amplifies runtime differences hidden by equal EPR cost, while increased communication concurrency can reduce the ability of EPR count to preserve runtime ordering. These results show that EPR count and execution time are related but distinct optimization objectives, and identify regimes in which explicit communication-aware runtime modeling is necessary.
\end{abstract}


\begin{IEEEkeywords}
distributed quantum computing,
quantum circuit partitioning,
communication-aware compilation,
quantum circuit scheduling,
EPR-pair consumption
\end{IEEEkeywords}


\section{Introduction}
\label{sec:introduction}

Distributed quantum computing (DQC) seeks to scale quantum computation by executing a quantum algorithm across multiple interconnected quantum processing units (QPUs). When logical qubits participating in the same two-qubit operation reside on different QPUs, the corresponding operation becomes non-local and requires inter-QPU communication, typically supported through shared entanglement, local operations, measurements, and classical communication \cite{caleffi_2024_distributed}. Circuit partitioning therefore determines which interactions become non-local and strongly influences the communication requirements of distributed execution.

A common objective in distributed compilation is to reduce the number of non-local interactions or the entanglement resources required to realize them. Several partitioning and compilation approaches use non-local gate count, teleportation cost, or EPR-pair consumption as measures of communication efficiency \cite{kaur_2025_optimized,burt_2025_entanglementefficient}. However, minimizing aggregate communication cost does not necessarily minimize runtime. Runtime also depends on where non-local operations occur within the circuit dependency structure, whether they can overlap, the latency of entanglement generation, and the availability of communication resources. These observations motivate the central question of this work: \emph{When is EPR-pair count a reliable proxy for distributed quantum
execution time?} 

Our results show that EPR count can be a strong runtime indicator in some
workload and communication regimes, while in others equal or lower EPR cost
can correspond to substantially different or even longer execution times. This work studies EPR-pair consumption as a runtime surrogate rather than proposing a new partitioning or compilation algorithm. The main contributions are as follows:

\begin{itemize}
    \item We characterize the relationship between EPR cost and runtime over alternative mappings of the same quantum circuit using rank correlation, equal-cost runtime ambiguity, rank reversals, and EPR-minimization regret.

    \item We exhaustively evaluate balanced static two-QPU mappings of QFT, QAOA, and CDKM circuits with $4$--$12$ logical qubits, yielding $1908$ candidate mappings.

    \item We characterize how proxy quality changes with EPR-generation latency and finite communication concurrency, showing that its reliability depends jointly on circuit structure and the communication operating regime.
\end{itemize}

\section{Related Work}
\label{sec:relatedwork}

Communication cost has long been a central objective in distributed quantum compilation. Researchers in \cite{cuomo_2023_optimizeda} already considered both entanglement-resource consumption and execution time, while the compiler framework in \cite{ferrari_2023_modular} combines partitioning and remote-operation scheduling with EPR-consumption and delay considerations. At the same time, increasingly sophisticated partitioning methods continue to target entanglement efficiency. For example, the temporally extended hypergraph formulation in \cite{burt_2025_entanglementefficient} accounts for multiple teleportation mechanisms while minimizing entanglement consumption. These works motivate an important distinction between minimizing communication resources and minimizing execution time.

A complementary line of research shows that runtime depends strongly on how communication demand interacts with the underlying execution resources. Network-aware scheduling under finite communication capacity is studied in \cite{pouryousef_2025_networkaware}, where the benefit of scheduling depends on available communication resources and the degree of entanglement parallelism exposed by the circuit. The subsequent benchmarking study in \cite{pouryousef_2026_benchmarking} further shows that end-to-end latency emerges from the interaction between workload structure and network conditions. More recently, architecture-aware compilation in \cite{kuo_2026_architectureaware} explicitly considers both EPR-pair consumption and communication runtime, reflecting the broader move toward joint computation--communication optimization.

Although prior work increasingly considers communication cost and execution time jointly, the validity of EPR count itself as a runtime surrogate has received less direct characterization. In particular, it remains important to understand whether EPR-based ordering of alternative mappings is preserved across different circuit structures and communication operating regimes. This work addresses this question through exhaustive mapping analysis while varying EPR-generation latency and communication concurrency, thereby identifying conditions under which EPR minimization is sufficient and conditions under which explicit runtime-aware modeling becomes necessary.


\section{System Model and Problem Formulation}
\label{sec:system_model}

For a fixed canonical circuit $\mathcal{C}$, let $\Pi(\mathcal{C})$ denote the set of candidate mappings of its logical qubits onto two QPUs. For each mapping $\pi\in\Pi(\mathcal{C})$, we evaluate its EPR communication cost $C_{\mathrm{EPR}}(\pi)$ and distributed execution time $T_{\mathrm{dist}}(\pi)$. We formalize the question introduced in Section~\ref{sec:introduction} as
\begin{equation}
    \operatorname*{arg\,min}_{\pi\in\Pi(\mathcal{C})}
    C_{\mathrm{EPR}}(\pi)
    \stackrel{?}{=}
    \operatorname*{arg\,min}_{\pi\in\Pi(\mathcal{C})}
    T_{\mathrm{dist}}(\pi).
    \label{eq:central_question}
\end{equation}
The remainder of this section defines the communication and execution models used to evaluate this relationship.

\subsection{Two-QPU Distributed Execution Model}
\label{subsec:two_qpu_model}

We consider two identical QPUs and a balanced, static logical-qubit mapping
\begin{equation}
    \pi:\{q_0,\ldots,q_{N_q-1}\}\rightarrow\{0,1\},
\end{equation}
where $N_q$ is the number of logical qubits and each QPU hosts $N_q/2$ qubits throughout execution. A two-qubit operation acting on $q_i$ and $q_j$ is local when $\pi(q_i)=\pi(q_j)$ and non-local when \( \pi(q_i)\neq\pi(q_j)\).

All single-qubit operations and local two-qubit operations are assigned the normalized duration \(T_{\mathrm{local}}=1.\) Let $T_{\mathrm{EPR}}$ denote the additional latency associated with generating the entanglement required for a non-local operation. Its cost relative to local computation is characterized by
\begin{equation}
    \rho_{\mathrm{EPR}}
    =
    \frac{T_{\mathrm{EPR}}}{T_{\mathrm{local}}}.
    \label{eq:rho_epr}
\end{equation}

Communication concurrency is represented by $N_{\mathrm{comm}}$, defined as the maximum number of non-local operations that may be active simultaneously. Each non-local operation occupies one communication slot for its full non-local-operation duration, whereas local operations consume no communication capacity. The baseline execution model assumes unconstrained communication concurrency, while finite $N_{\mathrm{comm}}$ is used to model communication contention.

\subsection{EPR Communication Cost}
\label{subsec:epr_cost}

Let $N_{\mathrm{NL}}(\pi)$ denote the number of canonical two-qubit operations whose operands are assigned to different QPUs under mapping $\pi$. In the baseline model, each such operation is assumed to consume one EPR pair, giving
\begin{equation}
    C_{\mathrm{EPR}}(\pi)
    =
    N_{\mathrm{NL}}(\pi).
    \label{eq:epr_cost}
\end{equation}

This controlled abstraction isolates aggregate entanglement demand from the temporal effects that determine execution time. It does not account for mechanisms that can alter EPR consumption, such as EPR reuse, pre-generation, or multi-gate teleportation.

In the remainder of this work, EPR cost refers specifically to EPR-pair count under the controlled assumption of one EPR pair per non-local canonical two-qubit operation.

\subsection{Execution-Time Model}
\label{subsec:execution_time}

Each canonical circuit is represented as a directed acyclic graph (DAG), $G=(V,E)$, where each node $v\in V$ represents one gate occurrence and each directed edge $(u,v)\in E$ represents an immediate precedence constraint. The dependency graph is fixed for a given circuit; a mapping $\pi$ changes which two-qubit operations are local or non-local, but does not alter the circuit dependencies.

Under mapping $\pi$, the duration of a circuit operation is
\begin{equation}
    d_{\pi}(v)=
    \begin{cases}
        T_{\mathrm{local}}, &
        \text{if $v$ is local},\\[1mm]
        T_{\mathrm{local}}+T_{\mathrm{EPR}}, &
        \text{if $v$ is non-local}.
    \end{cases}
    \label{eq:operation_duration}
\end{equation}

We first consider unconstrained communication concurrency. Each operation starts as soon as all of its predecessors have completed. Denoting its start and finish times by $s(v)$ and $f(v)$, respectively,
\begin{align}
    s(v)
    &=
    \max_{u\in\mathrm{Pred}(v)} f(u),
    \label{eq:start_time}\\
    f(v)
    &=
    s(v)+d_{\pi}(v),
    \label{eq:finish_time}
\end{align}
where $\mathrm{Pred}(v)$ denotes the set of predecessors of circuit operation $v$, and $s(v)=0$ for source nodes. The resulting distributed execution time is the circuit runtime,
\begin{equation}
    T_{\mathrm{dist}}(\pi)
    =
    \max_{v\in V} f(v).
    \label{eq:makespan}
\end{equation}

Finite communication concurrency extends this model by allowing a non-local operation to start only when both its circuit dependencies and a communication slot are available. Let
\begin{equation}
    r(v)
    =
    \max_{u\in\mathrm{Pred}(v)} f(u)
\end{equation}
denote the dependency-ready time of operation $v$, with $r(v)=0$ for source nodes. Local operations start at $s(v)=r(v)$. For a non-local operation, if $a_k$ denotes the availability time of its assigned communication slot, the start time becomes
\begin{equation}
    s(v)=\max\{r(v),a_k\}.
    \label{eq:finite_capacity_start}
\end{equation}
The finish time and runtime remain defined by \eqref{eq:finish_time} and \eqref{eq:makespan}, respectively.

This formulation separates aggregate EPR consumption from its temporal consequences. Two mappings with the same $C_{\mathrm{EPR}}$ may therefore produce different runtimes if their non-local operations occur at different positions in the dependency graph, admit different degrees of overlap, or experience different communication contention.


\section{Experimental Methodology}

\subsection{Benchmark Circuits and Candidate Mappings}
\label{subsec:benchmarks}

We evaluate three benchmark families with distinct interaction and dependency structures: the Quantum Fourier Transform (QFT), the Quantum Approximate Optimization Algorithm (QAOA), and the Cuccaro--Draper--Kutin--Moulton (CDKM) ripple-carry adder. QFT exhibits dense and increasingly global two-qubit interactions, QAOA contains graph-defined interactions repeated across algorithmic layers, whereas CDKM exhibits a strongly sequential carry-propagation structure. For each family, circuits with $N_q\in\{4,6,8,10,12\}$ logical qubits are considered. The QAOA instances use $p=2$ with seed $10$, while CDKM uses the full ripple-carry configuration.

The circuits are generated using the target-independent interface of MQT Bench~\cite{quetschlich_2023_mqt} to avoid assumptions about a particular processor topology or native gate set. Each benchmark is transformed into a common canonical representation before distributed mapping. Final measurements and barriers are removed, and the remaining circuit is transpiled with optimization level~0 into the logical basis
\begin{equation}
    \mathcal{B}_{\mathrm{can}}=\{u,cp\},
\end{equation}
where $u$ denotes a generic single-qubit operation and $cp$ a controlled-phase two-qubit operation. This basis serves only as a common logical abstraction and is not intended to represent the native gate set of a specific hardware platform.

Optimization is disabled both during benchmark generation and canonical transpilation. The objective is not to obtain a gate-count-optimal implementation, but to preserve a deterministic circuit representation across all candidate mappings. Hence, mappings of the same benchmark are evaluated against an identical logical workload, so differences in communication cost and execution time arise from the distributed mapping and execution model rather than from circuit recompilation.

Table~\ref{tab:canonical_benchmark_statistics} summarizes the resulting circuits, where $N_g$, $N_{1q}$, and $N_{2q}$ denote the total, single-qubit, and two-qubit gate counts, respectively, and $D$ denotes circuit depth.

\begin{table}[t]
\centering
\caption{Structural properties of the canonical benchmark circuits}
\small
\label{tab:canonical_benchmark_statistics}
\setlength{\tabcolsep}{8pt}
\begin{tabular}{lccccc}
\toprule
Benchmark & $N_q$ & $N_g$ & $N_{1q}$ & $N_{2q}$ & $D$ \\
\midrule
QFT-4   & 4  & 28  & 16  & 12 & 13 \\
QFT-6   & 6  & 48  & 24  & 24 & 17 \\
QFT-8   & 8  & 72  & 32  & 40 & 21 \\
QFT-10  & 10 & 100 & 40  & 60 & 25 \\
QFT-12  & 12 & 132 & 48  & 84 & 29 \\
\midrule
QAOA-4  & 4  & 24  & 20  & 4  & 11 \\
QAOA-6  & 6  & 60  & 46  & 14 & 17 \\
QAOA-8  & 8  & 108 & 80  & 28 & 37 \\
QAOA-10 & 10 & 186 & 134 & 52 & 49 \\
QAOA-12 & 12 & 252 & 180 & 72 & 59 \\
\midrule
CDKM-4  & 4  & 69  & 52  & 17 & 47 \\
CDKM-6  & 6  & 135 & 102 & 33 & 91 \\
CDKM-8  & 8  & 201 & 152 & 49 & 135 \\
CDKM-10 & 10 & 267 & 202 & 65 & 179 \\
CDKM-12 & 12 & 333 & 252 & 81 & 223 \\
\bottomrule
\end{tabular}
\end{table}

For each canonical circuit, we exhaustively enumerate balanced static mappings onto two identical QPUs according to the model in
Section~\ref{subsec:two_qpu_model}. Because exchanging the labels of the two QPUs produces an equivalent partition, $q_0$ is fixed to QPU~0 to remove this symmetry. The resulting number of unique mappings is
\begin{equation}
    N_{\mathrm{map}}(N_q)
    =
    \binom{N_q-1}{N_q/2-1}
    =
    \frac{1}{2}\binom{N_q}{N_q/2}.
    \label{eq:num_mappings}
\end{equation}
For $N_q\in\{4,6,8,10,12\}$, this gives $3$, $10$, $35$, $126$, and $462$ mappings per benchmark family, respectively. Across the three families, the study therefore evaluates $1908$ candidate mappings.

For every candidate, the canonical circuit remains unchanged while the logical-qubit grouping determines which two-qubit operations become non-local. Exhaustive enumeration ensures that subsequent EPR-cost and runtime comparisons cover the complete balanced two-QPU mapping space rather than the output of a particular partitioning heuristic.

\subsection{Scheduling and Validation}
\label{subsec:scheduler_validation}

For each canonical circuit, the dependency DAG defined in Section~\ref{subsec:execution_time} is constructed once and reused across all candidate mappings. Each mapping $\pi$ changes only the local or non-local classification of two-qubit operations and, consequently, their mapping-dependent durations; the underlying dependency structure remains unchanged.

Under unconstrained communication concurrency, the annotated DAG is evaluated using a deterministic earliest-start scheduler. Each operation begins as soon as all of its predecessors have completed, allowing dependency-independent operations to overlap. For finite $N_{\mathrm{comm}}$, the scheduler is extended to an earliest-feasible-start list scheduler, in which a non-local operation may begin only when both its dependencies are satisfied and a communication slot is available. This finite-capacity scheduler provides a controlled model of communication contention and is not intended to solve the globally optimal resource-constrained scheduling problem.

The implementation was validated against analytically tractable test circuits and the benchmark suite. DAG depths were checked against the corresponding Qiskit circuit depths, generated schedules were verified to satisfy all precedence constraints, and finite-capacity schedules exhibited the expected monotonic behavior with increasing communication concurrency.

\subsection{Evaluation Metrics}
\label{subsec:evaluation_metrics}

To assess the reliability of EPR cost as a runtime proxy, we consider complementary measures that capture ambiguity among equal-cost mappings, global ordering, pairwise ordering violations, and the optimization consequence of minimizing EPR consumption.

First, we investigate whether equal EPR costs can hide different runtimes. For mappings sharing the same EPR cost, the relative runtime spread is defined as
\begin{equation}
    S_{\mathrm{equal}}
    =
    \frac{T_{\max}-T_{\min}}{T_{\min}},
    \label{eq:equal_spread}
\end{equation}
where $T_{\min}$ and $T_{\max}$ are the minimum and maximum runtimes within the equal-cost group. A nonzero $S_{\mathrm{equal}}$ therefore captures runtime variation that is invisible to EPR count alone.

Next, we examine whether EPR cost preserves the overall runtime ordering of candidate mappings. For this, Spearman's rank correlation coefficient $\rho_s$ is computed between $C_{\mathrm{EPR}}(\pi)$ and $T_{\mathrm{dist}}(\pi)$ over all mappings of the same circuit. Values approaching $1$ indicate strong agreement between the EPR-cost and runtime orderings. The coefficient is undefined when either quantity is constant across all candidate mappings.

Then, we analyze whether lower EPR cost can correspond to higher runtime. A strict pairwise rank reversal occurs between mappings $\pi_i$ and $\pi_j$ when
\begin{equation}
    (C_i-C_j)(T_i-T_j)<0,
    \label{eq:rank_reversal}
\end{equation}
where $C_i=C_{\mathrm{EPR}}(\pi_i)$ and $T_i=T_{\mathrm{dist}}(\pi_i)$. The rank-reversal fraction $F_{\mathrm{rev}}$ is the fraction of strictly comparable mapping pairs satisfying \eqref{eq:rank_reversal}, excluding ties in either EPR cost or runtime.

Finally, we examine whether minimizing EPR cost recovers a runtime-optimal mapping. Let
\begin{equation}
    \mathcal{E}
    =
    \operatorname*{arg\,min}_{\pi\in\Pi(\mathcal{C})}
    C_{\mathrm{EPR}}(\pi),
    \qquad
    T^{*}
    =
    \min_{\pi\in\Pi(\mathcal{C})}
    T_{\mathrm{dist}}(\pi),
\end{equation}
where $\mathcal{E}$ is the set of minimum-EPR mappings and $T^{*}$ is the globally minimum execution time. If $T_{\mathcal{E}}^{\min}$ and $T_{\mathcal{E}}^{\max}$ denote the best and worst runtimes within $\mathcal{E}$, the corresponding regrets are
\begin{align}
    R_{\mathrm{EPR}}^{\mathrm{best}}
    &=
    \frac{T_{\mathcal{E}}^{\min}-T^{*}}{T^{*}},
    \label{eq:best_regret}\\
    R_{\mathrm{EPR}}^{\mathrm{worst}}
    &=
    \frac{T_{\mathcal{E}}^{\max}-T^{*}}{T^{*}}.
    \label{eq:worst_regret}
\end{align}
Thus, the regret values distinguish three cases. If $R_{\mathrm{EPR}}^{\mathrm{best}}=R_{\mathrm{EPR}}^{\mathrm{worst}}=0$, every minimum-EPR mapping is also runtime-optimal. If $R_{\mathrm{EPR}}^{\mathrm{best}}=0$ but $R_{\mathrm{EPR}}^{\mathrm{worst}}>0$, at least one minimum-EPR mapping is runtime-optimal, but EPR cost alone cannot identify which minimum-cost mapping to choose. Finally, if $R_{\mathrm{EPR}}^{\mathrm{best}}>0$, no minimum-EPR mapping achieves the globally minimum execution time.

\begin{figure*}[!t]
    \centering
    \includegraphics[width=\textwidth]{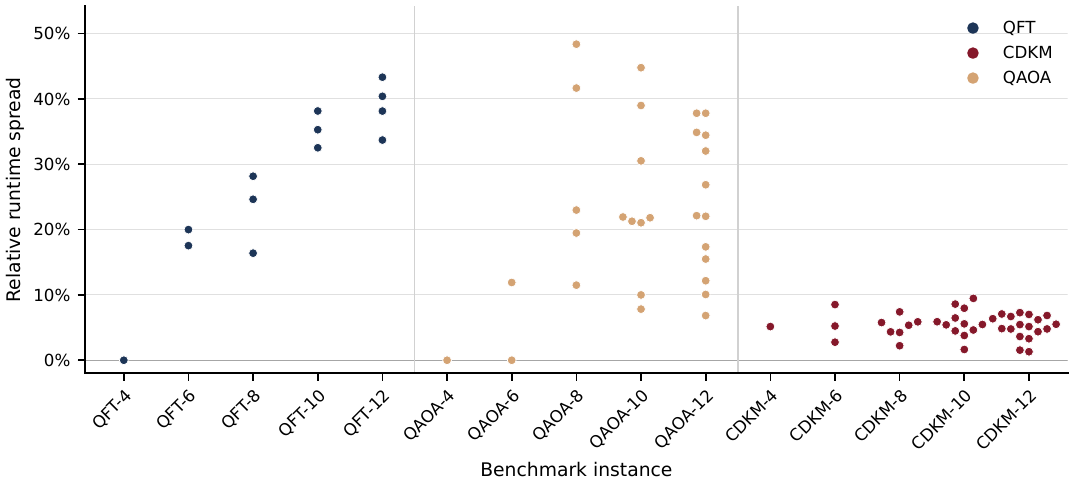}
    \caption{Relative runtime spread among mappings sharing the same EPR cost at $\rho_{\mathrm{EPR}}=5$. Each marker represents one EPR-cost group containing at least two mappings. A nonzero value indicates that mappings with identical EPR consumption produce different distributed execution times.}
    \label{fig:equal_epr_runtime_spread}
\end{figure*}


\section{Results}
\label{sec:results}

We first characterize the relationship between EPR cost and distributed execution time at the baseline operating point $\rho_{\mathrm{EPR}}=5$ under unconstrained communication concurrency. We then vary EPR-generation latency and communication concurrency to determine how the reliability of EPR cost changes across communication operating regimes.

\subsection{Baseline Relationship Between EPR Cost and Runtime}
\label{subsec:baseline_results}

The baseline experiment evaluates all $1908$ balanced two-QPU mappings of the $15$ benchmark instances. We first examine whether mappings with identical EPR cost can nevertheless exhibit different distributed execution times.

\subsubsection{Equal EPR Cost, Unequal Runtime}
\label{subsubsec:equal_epr_results}

Fig.~\ref{fig:equal_epr_runtime_spread} shows the relative runtime spread of every EPR-cost group containing at least two candidate mappings. Across the complete baseline dataset, $84$ such multi-candidate equal-EPR groups are observed. Of these, $81$ exhibit more than one execution time, corresponding to an ambiguity rate of $96.43\%$. Equal-EPR runtime ambiguity occurs in $13$ of the $15$ benchmark instances, showing that identical aggregate EPR consumption generally does not uniquely determine distributed execution time.

The magnitude of this ambiguity also varies substantially across circuit families and instances. The largest observed spread occurs for QAOA-8: five mappings with $C_{\mathrm{EPR}}=12$ produce execution times ranging  from $62$ to $92$, giving $S_{\mathrm{equal}}=48.39\%$. Large spreads are also observed for QAOA-10 and QFT-12, where the most pronounced equal-cost groups reach $44.79\%$ and $43.33\%$, respectively. Table~\ref{tab:equal_epr_largest_spreads} summarizes the five largest equal-EPR runtime spreads observed in the baseline experiment.

These results demonstrate that equal EPR count alone is insufficient to distinguish the execution times of mappings that incur the same communication cost.

\begin{table}[h]
    \centering
    \caption{Largest equal-EPR runtime spreads at
    $\rho_{\mathrm{EPR}}=5$}
    \label{tab:equal_epr_largest_spreads}
    \small
    \setlength{\tabcolsep}{3.2pt}
    \begin{tabular}{lccccc}
        \toprule
        Instance &
        $C_{\mathrm{EPR}}$ &
        Group size &
        $T_{\min}$ &
        $T_{\max}$ &
        $S_{\mathrm{equal}}$ \\
        \midrule
        QAOA-8  & 12 & 5   & 62  & 92  & 48.39\% \\
        QAOA-10 & 30 & 36  & 96  & 139 & 44.79\% \\
        QFT-12  & 42 & 180 & 90  & 129 & 43.33\% \\
        QAOA-8  & 16 & 9   & 72  & 102 & 41.67\% \\
        QFT-12  & 54 & 32  & 99  & 139 & 40.40\% \\
        \bottomrule
    \end{tabular}
\end{table}

\subsubsection{Global Ordering Between EPR Cost and Runtime}
\label{subsubsec:global_ordering}

\begin{figure}[!b]
    \centering
    \includegraphics[width=\columnwidth]{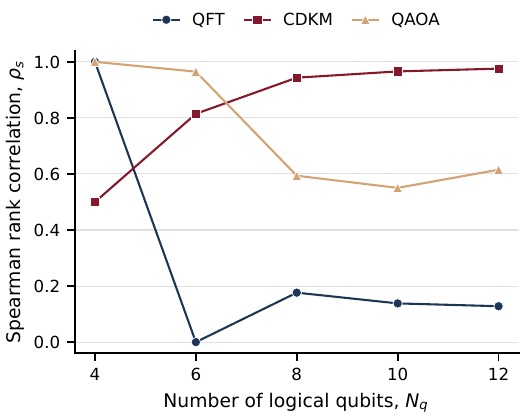}
    \caption{Spearman rank correlation between EPR cost and distributed execution time at $\rho_{\mathrm{EPR}}=5$. Values closer to $1$ indicate stronger agreement between the EPR-cost and runtime orderings of candidate mappings.}
    \label{fig:spearman_vs_qubits}
\end{figure}

We next examine whether EPR cost preserves the overall runtime ordering of candidate mappings. Fig.~\ref{fig:spearman_vs_qubits} reports Spearman's rank correlation coefficient $\rho_s$ between $C_{\mathrm{EPR}}$ and $T_{\mathrm{dist}}$ for each benchmark instance. The results reveal markedly different ordering behavior across the three circuit families.

For QFT, the ordering agreement becomes weak beyond the smallest instance. While QFT-4 gives $\rho_s=1$, the coefficient falls to $0$ for QFT-6 and remains between $0.128$ and $0.176$ for the 8--12-qubit instances. Thus, for the larger QFT circuits, ordering candidate mappings by EPR cost provides little agreement with their runtime ordering.

QAOA exhibits an intermediate behavior. The two smallest instances show strong agreement, with $\rho_s=1$ for QAOA-4 and $\rho_s=0.965$ for QAOA-6. For the larger instances, however, the correlation decreases to $0.594$, $0.550$, and $0.615$ for 8, 10, and 12 qubits, respectively. EPR cost therefore retains a positive association with runtime ordering for these instances, but does not reproduce that ordering closely.

In contrast, CDKM shows increasingly strong agreement as the circuit size grows. The correlation increases from $\rho_s=0.500$ at four qubits to $0.815$ at six qubits, and then reaches $0.943$, $0.966$, and $0.976$ for the 8-, 10-, and 12-qubit instances, respectively. For the larger CDKM circuits, EPR cost therefore provides a close approximation to the runtime ordering of the candidate mappings.

Overall, the baseline results show that the global ordering agreement between EPR cost and execution time is strongly dependent on the benchmark family and circuit size. EPR count can closely preserve runtime ordering in some instances, particularly the larger CDKM circuits, while providing only weak ordering information for others, most notably the larger QFT circuits.

\subsubsection{Strict Rank Reversals}
\label{subsubsec:rank_reversals}

\begin{figure}[h]
    \centering
    \includegraphics[width=\columnwidth]
    {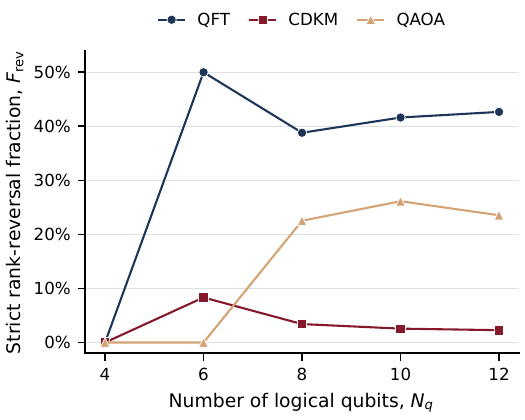}
    \caption{Strict rank-reversal fraction between EPR cost and distributed execution time at $\rho_{\mathrm{EPR}}=5$. A reversal occurs when the mapping with lower EPR cost has the longer execution time. Ties in either quantity are excluded.}
    \label{fig:rank_reversal_vs_qubits}
\end{figure}

We next examine whether the ordering induced by EPR cost can be strictly incorrect for individual mapping pairs. Fig.~\ref{fig:rank_reversal_vs_qubits} reports the strict rank-reversal fraction $F_{\mathrm{rev}}$, defined over candidate pairs that are not tied in either EPR cost or execution time.

The QFT instances exhibit the largest reversal fractions. No reversal is observed for QFT-4, but $F_{\mathrm{rev}}$ reaches $50.00\%$ for QFT-6 and remains high for the larger instances, with values of $38.81\%$, $41.63\%$, and $42.67\%$ for 8, 10, and 12 qubits, respectively. Thus, for a substantial fraction of strictly comparable QFT mapping pairs, the lower-EPR mapping is not the faster one.

QAOA exhibits fewer reversals than QFT but still shows a pronounced departure from EPR-based ordering for the larger instances. No strict reversals occur for QAOA-4 or QAOA-6, whereas the reversal fraction increases to $22.53\%$, $26.14\%$, and $23.54\%$ for QAOA-8, QAOA-10, and QAOA-12, respectively.

CDKM again displays a substantially stronger correspondence between EPR cost and execution time. Its reversal fraction is $0$ for CDKM-4, $8.33\%$ for CDKM-6, and decreases to $3.41\%$, $2.57\%$, and $2.28\%$ for the 8-, 10-, and 12-qubit instances. Consequently, strict ordering violations become relatively rare for the larger CDKM circuits.

These results strengthen the conclusion obtained from the global rank-correlation analysis. Weak correlation is not caused only by ties or small changes in ordering: for QFT and the larger QAOA instances, lower EPR cost can directly correspond to longer execution time. However, the frequency of such reversals remains strongly dependent on the benchmark family.

\subsubsection{Runtime Consequences of EPR-Cost Minimization}
\label{subsubsec:epr_regret}

Finally, we examine whether minimizing EPR cost recovers a runtime-optimal mapping. Table~\ref{tab:epr_regret_baseline} reports the best- and worst-case runtime regrets within the set of minimum-EPR mappings. The two quantities distinguish whether the runtime optimum is contained in the minimum-EPR set and, when multiple minimum-cost mappings exist, how much their execution times may differ.

For QFT, the best-case regret is zero for every instance, indicating that the minimum-EPR set always contains at least one runtime-optimal mapping. However, the worst-case regret becomes substantial for the larger instances, reaching $20.0\%$, $16.4\%$, $38.2\%$, and $33.7\%$ for QFT-6 through QFT-12, respectively. Thus, although EPR minimization can recover the runtime optimum for these instances, EPR cost alone cannot reliably distinguish among tied minimum-cost mappings.

QAOA exhibits both forms of mismatch. QAOA-4 and QAOA-6 have zero best- and worst-case regret, whereas QAOA-8 retains zero best-case regret but exhibits the largest worst-case regret in the baseline study, $48.4\%$. More importantly, QAOA-10 and QAOA-12 have positive best-case regrets of $6.4\%$ and $1.7\%$, respectively. For these two instances, no minimum-EPR mapping achieves the globally minimum execution time.

CDKM shows the strongest agreement between EPR minimization and runtime minimization. CDKM-4 has zero best-case regret and a modest worst-case regret of $5.2\%$, while all larger CDKM instances have zero best- and worst-case regret. Hence, for CDKM-6 through CDKM-12, every minimum-EPR mapping is also runtime-optimal under the baseline model.

Overall, the optimization consequence of using EPR cost is therefore workload dependent. In some instances, minimizing EPR cost is sufficient to recover the runtime optimum; in others, ties among minimum-EPR mappings leave substantial runtime ambiguity, while in still others the minimum-EPR set itself excludes the runtime-optimal mapping.

\begin{table}[h]
    \centering
    \caption{Best- and worst-case runtime regret among minimum-EPR
    mappings at $\rho_{\mathrm{EPR}}=5$. Each entry reports
    $(R_{\mathrm{EPR}}^{\mathrm{best}},
    R_{\mathrm{EPR}}^{\mathrm{worst}})$ in percent.}
    \label{tab:epr_regret_baseline}
    \small
    \setlength{\tabcolsep}{3.2pt}
    \begin{tabular}{lccccc}
        \toprule
        & \multicolumn{5}{c}{$N_q$} \\
        \cmidrule(lr){2-6}
        Family & 4 & 6 & 8 & 10 & 12 \\
        \midrule
        QFT
        & $(0,0)$
        & $(0,20.0)$
        & $(0,16.4)$
        & $(0,38.2)$
        & $(0,33.7)$ \\

        QAOA
        & $(0,0)$
        & $(0,0)$
        & $(0,48.4)$
        & $(6.4,17.0)$
        & $(1.7,8.7)$ \\

        CDKM
        & $(0,5.2)$
        & $(0,0)$
        & $(0,0)$
        & $(0,0)$
        & $(0,0)$ \\
        \bottomrule
    \end{tabular}
\end{table}

\begin{figure*}[!t]
    \centering
    \includegraphics[width=\textwidth]{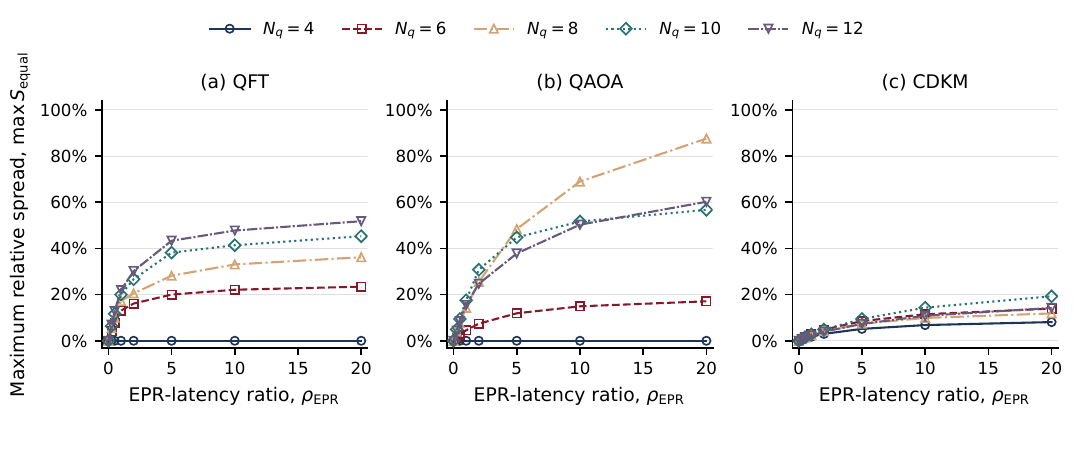}
    \caption{Maximum relative runtime spread $S_{\mathrm{equal}}$ among mappings with equal EPR cost as a function of normalized EPR-generation latency. Subfigures (a)--(c) correspond to QFT, QAOA, and CDKM, respectively; curves denote circuit size $N_q$.}
    \label{fig:latency_equal_epr}
\end{figure*}

\begin{figure*}[!t]
    \centering
    \includegraphics[width=\textwidth]{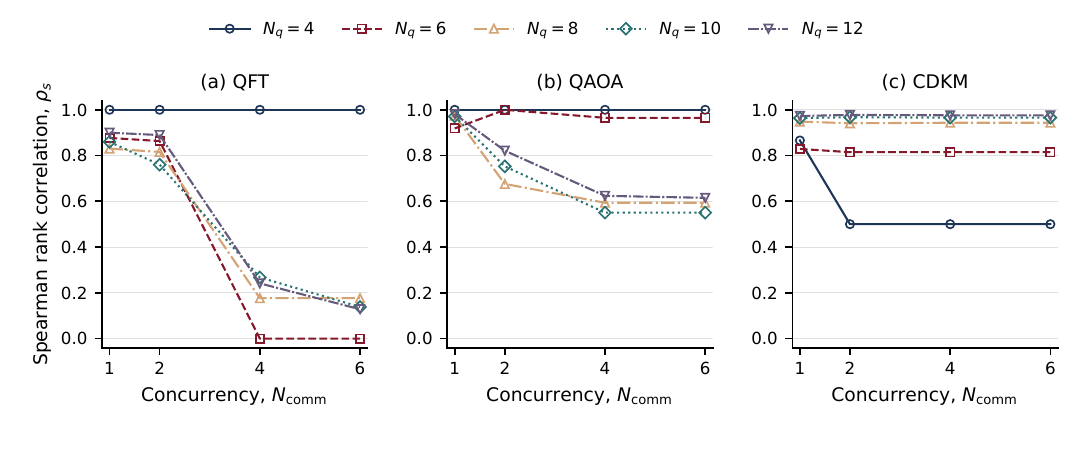}
    \caption{Spearman rank correlation $\rho_s$ between EPR cost and distributed execution time as a function of communication concurrency $N_{\mathrm{comm}}$ at $\rho_{\mathrm{EPR}}=5$. Subfigures (a)--(c) correspond to QFT, QAOA, and CDKM, respectively;
    curves denote circuit size $N_q$.}
    \label{fig:spearman_communication_capacity}
\end{figure*}

\subsection{Effect of EPR-Generation Latency}
\label{subsec:epr_latency_results}

We next examine how EPR-generation latency affects the discrepancy between EPR cost and distributed execution time. Keeping the benchmark circuits, candidate mappings, and unconstrained communication model fixed, we vary the normalized EPR-generation latency over $\rho_{\mathrm{EPR}}\in\{0,0.25,0.5,1,2,5,10,20\}$. At each operating point, all $1908$ candidate mappings are reevaluated. The case $\rho_{\mathrm{EPR}}=0$ serves as a zero-communication-penalty reference, while increasing $\rho_{\mathrm{EPR}}$ progressively raises the runtime contribution of each non-local operation.

To isolate the effect of latency from EPR count, we focus on mappings with identical EPR cost and measure their relative runtime spread. This directly captures how increasing EPR-generation latency amplifies runtime differences that are invisible to EPR count.

Fig.~\ref{fig:latency_equal_epr} reports, for each benchmark instance, the maximum relative runtime spread $S_{\mathrm{equal}}$ among equal-EPR mapping groups. At $\rho_{\mathrm{EPR}}=0$, the spread is zero for every instance because local and non-local operations have identical durations. Once communication incurs a nonzero latency penalty, however, mappings with the same EPR count can exhibit different execution times, and this discrepancy generally grows as the EPR-generation latency penalty increases.

The effect is strongest for QFT and QAOA. For QFT-12, the maximum equal-EPR spread increases from $7.0\%$ at $\rho_{\mathrm{EPR}}=0.25$ to $43.3\%$ at the baseline operating point $\rho_{\mathrm{EPR}}=5$, reaching $51.8\%$ at $\rho_{\mathrm{EPR}}=20$. QFT-10 follows a similar trend, increasing from $6.4\%$ to $45.3\%$ over the sampled positive-latency range.

The largest growth occurs for QAOA-8, whose maximum spread rises from $3.9\%$ at $\rho_{\mathrm{EPR}}=0.25$ to $48.4\%$ at $\rho_{\mathrm{EPR}}=5$ and $87.6\%$ at $\rho_{\mathrm{EPR}}=20$. QAOA-10 and QAOA-12 also show substantial growth, reaching maximum spreads of $56.7\%$ and $60.2\%$, respectively, at the largest sampled latency. CDKM exhibits the same qualitative trend but with considerably smaller magnitudes; its largest observed spread is $19.3\%$, for CDKM-10 at $\rho_{\mathrm{EPR}}=20$.

These results show that EPR-generation latency can amplify runtime differences that are completely invisible to EPR count. Importantly, the magnitude of this amplification is workload dependent: equal-EPR mappings remain comparatively similar in runtime for CDKM, whereas QFT and especially QAOA can diverge substantially as the EPR-generation latency penalty increases. 

\subsection{Effect of Communication Concurrency}
\label{subsec:communication_capacity_results}

We finally examine how communication concurrency affects the reliability of EPR cost as a runtime proxy. Keeping $\rho_{\mathrm{EPR}}=5$ fixed, we limit the number of simultaneous non-local operations to $N_{\mathrm{comm}}\in\{1,2,4,6\}$. Since changing communication concurrency primarily affects the amount of overlap available between non-local operations, we use the Spearman rank correlation $\rho_s$ to measure how well EPR cost preserves the global runtime ordering of candidate mappings.

Fig.~\ref{fig:spearman_communication_capacity} shows that the effect of communication concurrency is strongly workload dependent. The largest change occurs for QFT. Under highly serialized communication, EPR cost provides a comparatively strong ordering of candidate runtimes; for example, QFT-12 has $\rho_s\approx0.90$ at $N_{\mathrm{comm}}=1$. As additional communication concurrency becomes available, the correlation deteriorates substantially, reaching approximately $0.13$ at $N_{\mathrm{comm}}=6$. QFT-6 exhibits an even more pronounced change, with its correlation falling to zero for the higher-concurrency settings.

QAOA shows the same general tendency for its larger instances, although the degradation is less severe. QAOA-8, QAOA-10, and QAOA-12 all exhibit lower rank correlation as communication concurrency increases, whereas the smaller QAOA instances remain strongly correlated. In contrast, CDKM is comparatively insensitive to additional concurrency: apart from CDKM-4, the correlations remain high and nearly unchanged across the tested concurrency settings.

These results indicate that stronger communication serialization can make EPR count a better predictor of execution time. When non-local operations are forced to execute more sequentially, makespan becomes more strongly related to their total number. As communication concurrency increases, non-local operations can overlap, making runtime increasingly dependent on their temporal placement, dependencies, and critical-path structure.

\subsection{Implications for Communication-Aware DQC Compilation}

The combined results indicate that EPR count should not be treated as a universally architecture-independent optimization objective. Its reliability depends jointly on circuit dependency structure, EPR-generation latency, and the amount of communication concurrency available during execution. Consequently, the suitability of an EPR-based partitioning objective should be evaluated together with the execution architecture for which the circuit is being compiled.

For workloads and operating regimes in which EPR count closely preserves runtime ordering, EPR minimization provides a simple and effective communication-aware objective. In regimes with substantial runtime ambiguity or frequent ordering violations, however, compilation should incorporate temporal information such as the placement of non-local operations in the dependency graph, critical-path effects, and the communication resources available for overlapping non-local operations.

These observations are particularly relevant to emerging modular quantum processors and quantum data-center architectures, where interconnect latency and the number of simultaneously available communication channels are expected to evolve alongside processor scale. Changes in these architectural capabilities can alter which circuit mappings are preferable even when their EPR costs remain unchanged. The methodology developed here can therefore be used to assess whether EPR-based objectives remain adequate as distributed quantum architectures and compilation strategies evolve.

\section{Conclusion}
\label{sec:conclusion}

This work examined when EPR-pair count provides a reliable surrogate for distributed quantum execution time. Rather than proposing a new partitioning algorithm, we evaluated the relationship between EPR cost and runtime directly by exhaustively considering balanced two-QPU mappings of QFT, QAOA, and CDKM benchmark circuits. The results show that minimizing aggregate EPR consumption and minimizing execution time are related but distinct objectives. Mappings with identical EPR cost can exhibit substantially different runtimes, lower-EPR mappings can execute more slowly than higher-EPR alternatives, and in some instances the minimum-EPR set does not contain the runtime-optimal mapping.

The results further show that the reliability of EPR count is not an intrinsic property of the cost metric alone, but depends jointly on the workload and the communication architecture. Under the evaluated model, EPR count can provide a useful communication-cost objective in some regimes, while other regimes require explicit consideration of communication timing and execution dependencies. This distinction is particularly relevant to communication-aware compilation for modular and networked quantum processors, where evolving interconnect latency and concurrency can change the relationship between entanglement consumption and execution time.

Future work can extend this characterization to larger multi-QPU systems, more detailed entanglement-generation models, and compilation mechanisms such as multi-gate teleportation.

\section*{Acknowledgments}
Fatih E. Bilgen’s work was supported by the TUBITAK BIDEB 2213-A Scholarship.


\bibliographystyle{IEEEtran}
\bibliography{references}

\end{document}